\documentclass[twocolumn,preprintnumbers,amsmath,amssymb]{revtex4}

\usepackage{graphicx}% Include figure files
\usepackage{bm}% bold math

\newcommand{\be}{\begin{equation}}
\newcommand{\ee}{\end{equation}}
\newcommand{\bea}{\begin{eqnarray}}
\newcommand{\eea}{\end{eqnarray}}

\newcommand{\Ptwo}{{\mathcal P}_2}
\newcommand{\bi}{\begin{itemize}}
\newcommand{\ei}{\end{itemize}}
\newcommand{\thint}{\widetilde{d\theta}}

\newcommand{\angint}{\int\!\thint\,}
\newcommand{\totint}{\int\!du\,\thint\,}
\newcommand{\fex}{f_{\rm{ex}}}

\newcommand{\fid}{f_{\rm{id}}}
\newcommand{\na}{n_a}

\newcommand{\xa}{x_a}
\newcommand{\xb}{x_b}
\newcommand{\parent}{n_0(u)}

\newcommand{\half}{\frac{1}{2}}
\newcommand{\eq}[1]{~(\ref{#1})}
\newcommand{\m}{\bar{S}}
\newcommand{\ma}{\m_a}
\newcommand{\mb}{\m_b}

\begin{document}

\title{Nematic-nematic demixing in polydisperse thermotropic liquid
crystals}

\author{Peter Sollich}
\email{peter.sollich@kcl.ac.uk}
\affiliation{Department of Mathematics, King's College London,
Strand, London WC2R 2LS, U.K.}

\begin{abstract}
We consider the effects of polydispersity on isotropic-nematic phase
equilibria in thermotropic liquid crystals, using a Maier-Saupe theory
with factorized interactions. A sufficient spread ($\approx 50\%$) in
the interaction strengths of the particles leads to phase separation
into two or more nematic phases, which can in addition coexist with an
isotropic phase. The isotropic-nematic coexistence region widens
dramatically as polydispersity is increased, leading to re-entrant
isotropic-nematic phase separation in some regions of the phase
diagram. We show that similar phenomena will occur also for
non-factorized interactions as long as the interaction strength
between any two particle species is lower than the mean of the
intra-species interactions.
\end{abstract}

\maketitle

\section{Introduction}

Solutions of anisotropic colloids or stiff polymers can exhibit phase
transitions from isotropic to orientationally ordered nematic states
as temperature is lowered or density is increased. Such systems are
often polydisperse, containing e.g.\ colloid particles with an
effectively continuous range of lengths and diameters or polymers with
a distribution of chain lengths. Such polydispersity has
experimentally measurable consequences. For example, in polydisperse
colloidal rods with nearly hard interactions, demixing into two
nematic (N) phases and coexistence of two nematics and an isotropic
(I) phase can occur~\cite{BuiLek93,VanVanLek96}. This is to be
contrasted with the hypothetical case where all colloidal particles
are identical, i.e.\ the solution is monodisperse; there at most two
phases (I--N) can coexist.

For the theoretical description of phase transitions to nematically
ordered states there are two broad classes of models. The first
focusses on short-range repulsions between particles and is therefore
appropriate for {\em lyotropic} liquid crystals whose phase changes
are controlled primarily by density. It contains in particular the
Onsager model of hard rods, whose second-virial truncation becomes
exact in the limit of long thin rods~\cite{Onsager49}, and the Flory
lattice model~\cite{Flory56b}. The second class of models concentrates
on the anisotropy of longer-range attractive interactions. It contains
in particular the Maier-Saupe model, a mean-field theory which expands
the orientational interaction dependence in low-order Legendre
polynomials~\cite{MaiSau58,HumJamLuc72}. Models of this type are
useful for {\em thermotropics}, where temperature is the dominant
control parameter.  Theories combining these two approaches have also
been developed, see the review~\cite{Singh00}, and extended to include
e.g. particle flexibility~\cite{KhoSem85}.

For {\em lyotropics}, binary (and to lesser extent ternary) mixtures
have been thoroughly investigated theoretically using the Onsager and
Flory models~\cite{AbeFlo78,LekCouVanDeb84,%
OdiLek85,BirKolPry88,VroLek93,VanMul96,Hemmer99,Hemmer00,VarSza00,VarGalJac03},
and phase separation into two nematic (N--N) phases and three-phase
I--N--N coexistence predicted for sufficiently disparate rod sizes.
Early work on binary {\em thermotropic} mixtures ignored the
possibility of fractionation, i.e.\ unequal partitioning of different
particle species across coexisting phases. Within this approximation a
mixture behaves like a monodisperse system, with no I--N coexistence
region~\cite{HumJamLuc71,HumLuc73}. Later studies that accounted for
fractionation effects did find I--N~\cite{HumLuc76}
%
%\cite{DavDew54,DavDew55} also saw none experimentally
%
and N--N~\cite{PalBerDebDun84} coexistence in bidisperse Maier-Saupe
theory. We note as an aside that similar effects have also been seen
in theories of mixtures of semi-flexible or stiff polymers; see
e.g.~\cite{KhoSem85,MaiSix89,HolSch92,ChiKyu95}. However, there a
driving force for demixing even between isotropic phases is normally
present, e.g.\ because of unfavourable interactions between unlike
species. N--N demixing then occurs rather trivially when the nematic
order prevalent at low temperatures is `superimposed' on this
isotropic demixing transition.

\iffalse
Comment on N-N in LC polymers, where it's nothing to do with
polydispersity, just ordinary demixing that would also take place for
isotropic phases, decorated by nematic ordering.
[\cite{ChiKyu95} - bidisperse Flory-Huggins plus Maier-Saupe for
semi-flexible polymers; get N-N but really a consequence of
Flory-Huggins. They also allow for non-factorized interactions (their
c is 1 for the factorized case). \cite{HolSch92} use fancier
technology but for rigid-rigid mixtures get similar phase diagrams;
also \cite{MaiSix89}. (also for single species in\cite{KhoSem85},
where N-N can again occur, but largely a Parsons-Onsager theory with
added Flory-Huggins attraction)]
\fi

%Some experimental evidence for this: \cite{CasVeyFin82} - mixture of
%nematic sidechain polymer and small LC molecule. They see a fairly
%wide N-N coexistence region, but it looks like this may be driven by
%Flory-Huggin type demixing in the ordered region.

Theoretical studies of genuinely polydisperse systems have been rather
rarer. For distributions of the polydisperse attribute (particle
length, diameter etc) with two narrow peaks one of course expects
phase behaviour similar to the bidisperse
case~\cite{ItoTer84,ItoTer84b}, but it is not at all obvious what
happens for unimodal (single-peaked) distributions.  For {\em
lyotropics}, work on the polydisperse Flory
model~\cite{FloAbe78,FloFro78,FroFlo78,MosWil82} focussed on general
features~\cite{Sollich02} such as fractionation and the widening of
the I--N coexistence region, but did not investigate the possibility of
N--N demixing. For the Onsager model, primarily polydispersity in
particle lengths has been studied. For a narrow distribution of
lengths, perturbative calculations show that the I--N coexistence
region again widens~\cite{Chen94}. Recent
studies~\cite{WenVro03,SpeSol03b} for wider length distributions have
confirmed this and provided evidence for a region of I--N--N
coexistence in the phase diagram. However, the existence of N--N
demixing at higher density remains an open question; in simplified
Onsager theories N--N coexistence is found only in a limited density
range~\cite{SpeSol02,SpeSol03a} or not at
all~\cite{ClaCueSeaSolSpe00}.

For polydisperse {\em thermotropics}, Sluckin used Maier-Saupe theory
to predict the opening up of an I--N coexistence region in slightly
polydisperse systems~\cite{Sluckin89}. This was based on a
perturbation theory around the monodisperse case, which did not allow
for a study of possible N--N demixing. Semenov obtained similar
results for thermotropic semi-flexible polymers with small
polydispersity in the chain lengths~\cite{Semenov93}.

Our aim in this paper is to find out whether a unimodal distribution
of the polydisperse attribute is enough to cause N--N demixing and the
associated I--N--N coexistence in polydisperse thermotropics. We use
the simplest Maier-Saupe theory for this purpose, with factorized
interactions. As explained in the next section, this gives a
truncatable free energy for which phase equilibria can be calculated
relatively efficiently. The resulting phase diagrams are presented in
Sec.~\ref{sec:results}, and wider implications and possible
generalizations are discussed in Sec.~\ref{sec:conclusions}.

%\cite{BuiLek93} has rather polydisperse rods (length around 50\%,
%diameter 25\%) and they do see N-N, though primarily hard interaction.
%
%could argue that attractive interactions should reinforce N-N
%separation observed there?
%
%Aim: investigate possibility of N-N demixing driven by unimodal
%polydispersity, with simplest thermotropic (Maier-Saupe) model.

\iffalse
Probably leave out: [Can also get N-N in monodisperse systems from
combination of isotropic and anisotropic interactions, see
\cite{TeiTel95} - weird single-species model with isotropic hard
repulsion, plus soft attractive/repulsive isotropic interaction $\sim
r^{-6}$, plus attractive Maier-Saupe term (also $\sim r^{-6}$). If the
soft isotropic interaction is repulsive, they get N-N coexistence in a
narrow T-range. Not very transparent physically.  \cite{Teixeira97}
also sees such effects, though for a different interaction potential.]
\fi

\section{Method}

We consider a system with liquid crystal particles having a continuous
range of values of some polydisperse attribute $l$. This could be
particle length, for example, or polarizability in the case of van der
Waals interactions; for definiteness we refer to length below. The
state of a single phase of a system is then described by a
distribution $n(l,\Omega)$, defined so that
$n(l,\Omega)\,dl\,d\Omega/4\pi$ gives the fraction of particles with
length in an interval $dl$ around $l$, and orientations $\Omega$ in a
solid angle $d\Omega$. This distribution is the natural extension for
a polydisperse system of the usual orientational distribution
$P(\Omega)$ used for monodisperse systems of rods~\cite{VroLek92}. The
rod orientation $\Omega$ can be parameterized in terms of the angle
$\theta$ with the nematic axis and an azimuthal angle $\varphi$; due
to the cylindrical symmetry of the nematic phase $n(l,\Omega)$ is
independent of $\varphi$. Using $d\Omega = d\cos\theta\ d\varphi$ the
length distribution, obtained by integrating over orientations, is
therefore
\begin{equation}\label{eq:normalization_ang}
n(l)=\frac{1}{4\pi}\int\! d\Omega\ \rho(l,\Omega)=\angint
n(l,\theta)
\end{equation}
where $\angint \ldots = (1/2)\int_{-1}^1
d\cos\theta\ \ldots$\ . The orientational distribution of the particles
can be factored out from $n(l,\theta)$ as~\cite{ClaCueSeaSolSpe00}
\begin{equation}
n(l,\theta)=
n(l)P_l(\theta)
\label{eq:decomposition_rho}
\end{equation}
where $P_l(\theta)$ represents the probability of finding a rod of
given length $l$ in orientation $\Omega=(\theta,\varphi)$ and is
normalized according to $\angint P_l(\theta)=1$. In the isotropic
phase, one has $P_l(\theta)\equiv 1$ and $n(l,\theta)=n(l)$.

%%%%%%%%%%%%%%%%%%%%%%%%%%%%%%%%%%%%%%%%%%%%%%%%%%%%%%%%%%%%%%%%%%%%%%
%State free energy following Sluckin, use l for polydisp attribute -
%take from P2 paper. Factorize interactions and switch to
%u. 
%%%%%%%%%%%%%%%%%%%%%%%%%%%%%%%%%%%%%%%%%%%%%%%%%%%%%%%%%%%%%%%%%%%%%%

In the above notation the free energy per particle of polydisperse
Maier-Saupe theory~\cite{Sluckin89} can be written as $f=\fid+\fex$,
where the ideal part $\fid$ is the free energy of an ideal mixture
while the excess part is
\be
\fex = - \half \int\! dl\,dl'\, n(l) n(l') u(l,l') S(l) S(l')
\ee
Here $S(l) = \angint P_l(\theta) \Ptwo(\cos\theta)$ is the
orientational order parameter of particles with length $l$, defined as
the average of the second Legendre polynomial
$\Ptwo(\cos\theta)=\frac{1}{2}(3\cos^2\theta-1)$. The coefficients
$u(l,l')$ determine the strength of the attractive interaction
favouring nematic ordering and depend on the particle lengths $l$ and
$l'$. We note that the description of the system in terms of a
normalized distribution over lengths and orientations rather than a
density distribution contains the implicit assumption that density
variations with temperature or between coexisting phases are weak and
can be neglected. Accordingly, we will not concern ourselves with the
density dependence of the interaction strengths $u(l,l')$. Maier and
Saupe~\cite{MaiSau58} assumed this to be quadratic, but
Cotter~\cite{Cotter77c} later showed that a thermodynamically
consistent derivation of Maier-Saupe theory as a mean-field
approximation requires a linear density scaling.

%%%%%%%%%%%%%%%%%%%%%%%%%%%%%%%%%%%%%%%%%%%%%%%%%%%%%%%%%%%%%%%%%%%%%%
%State equilibrium conditions (equal chemical potentials), write
%down equations for coexisting distributions.
%%%%%%%%%%%%%%%%%%%%%%%%%%%%%%%%%%%%%%%%%%%%%%%%%%%%%%%%%%%%%%%%%%%%%%
%
In the following we make the common assumption that the interaction
strengths factorize according to $u(l,l')=[u(l,l)u(l',l')]^{1/2}$. It
is then sensible to switch from $l$ to $u\equiv u(l,l)$ as the
polydisperse attribute, giving for the ideal and excess free energies
($k_{\rm B}=1$) 
\bea
\fid &=& T\totint n(u,\theta)\ln n(u,\theta)\\
&=&T\int\! du\, n(u)\left[\ln n(u) + \angint P_u(\theta)\ln
P_u(\theta)\right]\\
\fex &=& - \half \int\! du\,du'\, n(u) n(u') \sqrt{uu'} S(u) S(u')
\eea
With our factorized interactions the underlying nature of the
polydisperse attribute has become irrelevant; polydispersity manifests
itself only through the spread in the interaction strengths $u$.

For a given distribution $n(u)$, the orientational distributions are
obtained by minimization of the free energy with respect to the
orientational distributions $P_u(\theta)$. Bearing in mind that these
are normalized, one finds
\bea
\label{eq:P_l}
P_u(\theta) &=& z^{-1}(u)\exp\left(\beta\m\sqrt{u}\Ptwo\right) \\
z(u) &=& \angint \exp\left(\beta\m\sqrt{u}\Ptwo\right)
\label{eq:z}
\eea
where we have abbreviated $\beta=1/T$ and $\Ptwo\equiv
\Ptwo(\cos\theta)$ and $z(u)$ is the normalizing partition function
for $P_u(\theta)$. The order parameter that appears
in~(\ref{eq:P_l},\ref{eq:z}) is
\be
\m = \int\! du\, n(u) \sqrt{u} S(u) = \totint n(u,\theta)\sqrt{u}\Ptwo
\ee
and determines the excess free energy $\fex = -\m^2/2$. It obeys the
self-consistency equation
\be
\label{eq:rho2_selfconsis}
\m = \totint n(u) 
\frac{\sqrt{u} 
\Ptwo \exp\left(\beta\m \sqrt{u} \Ptwo\right)}
{\angint \exp\left(\beta\m \sqrt{u} \Ptwo\right)}
\ee
This always has the trivial solution $\m=0$ corresponding to an
isotropic phase; nematic phases are characterized by $\m>0$ and become
physically relevant where they have a lower free energy than the
isotropic solution. 

To calculate phase equilibria we need the chemical potentials
$\mu(u)$, which need to be equal in coexisting phases. Denote $N(u)$
the particle number distribution, so that the total particle number is
$N=\int\! du\,N(u)$, and $n(u)=N(u)/N$. Then $\mu(u) = \delta F/\delta
N(u)$, where the extensive free energy is $F=Nf[N(u)/N]$. This gives
\be
\mu(u) = \frac{\delta f}{\delta n(u)} + f - \int
du'\,n(u')\frac{\delta f}{\delta n(u')}
\ee
In evaluating the derivatives $\delta f/\delta n(u)$ we do not need to
take into account the variation of the $P_u(\theta)$ because they are
chosen to minimize $f$. A short calculation then gives $\delta
f/\delta n(u)=T\ln(n(u)/z(u))+T$ and so
\be
\mu(u) = T\ln\frac{n(u)}{z(u)} + \half \m^2
\ee
If the system separates into $P$ phases $a=1,\ldots,P$,
their $u$-distributions $\na(u)$ can therefore be written as
\be
\na(u) = R(u) z_a(u) \exp(-\beta\ma^2/2)
\ee
The common factor $R(u)=\exp[\beta \mu(u)]$ follows from the
requirement of particle conservation: if phase $a$ contains a fraction
$\xa$ of all particles, then $\sum_a \xa \na(u) = \parent$, where
$\parent$ is the overall or ``parent'' distribution of $u$. Using
also\eq{eq:P_l}, the joint distribution over interaction strengths and
orientations in each phase therefore takes the form
\be
\na(u,\theta) = 
\frac{\parent \exp\left(\beta\ma \sqrt{u} \Ptwo - \beta \ma^2/2\right)}
{\sum_b \xb\angint \exp\left(\beta\mb \sqrt{u} \Ptwo - \beta \mb^2/2\right)}
\ee
A phase split involving $P$ phases is therefore characterized by the
$2P$ variables $\ma$ and $\xa$. These are determined by as many
conditions, namely 
\bea
1   &=& \totint \na(u,\theta)
\label{na-norm}
\\
\ma &=& \totint \na(u,\theta)\sqrt{u}\Ptwo
\eea
The relations\eq{na-norm} automatically guarantee that $\sum_a \xa=1$
because the parent $\parent$ is normalized by assumption.

%%%%%%%%%%%%%%%%%%%%%%%%%%%%%%%%%%%%%%%%%%%%%%%%%%%%%%%%%%%%%%%%%%%%%%
%Explain numerics:
%Truncatable theory, adapt from ~\cite{SpeSol02,SpeSol03a}.
%%%%%%%%%%%%%%%%%%%%%%%%%%%%%%%%%%%%%%%%%%%%%%%%%%%%%%%%%%%%%%%%%%%%%%

In our numerical work, we exploit the fact that the excess free energy
$\fex = -\m^2/2$ is truncatable~\cite{SolWarCat01}, i.e.\ it involves
only a single moment $\m$ of the distribution $n(u,\theta)$. Phase
equilibria can then efficiently be found using the moment free energy
(MFE) method~\cite{SolCat98,Warren98,SolWarCat01}. This constructs a
MFE which depends only on $\m$, and from which phase coexistence can
be predicted in the standard manner. In this
simplest form the MFE gives only the onset of phase separation
exactly. However, by including additional moments, defined by
adaptively chosen weight functions, increasingly accurate solutions
are obtained. Using these as initial points, we are then able to find
the, for our free energy, exact solutions of the phase equilibrium
equations~\cite{ClaCueSeaSolSpe00,SpeSol02,SpeSol03a}. We omit the
details of the implementation because they are similar to our study of
an approximate Onsager model~\cite{SpeSol02,SpeSol03a}. Differences
arise primarily because, in the Maier-Saupe model, we are describing
the system in terms of normalized distributions, as would be
appropriate in other systems for the case of fixed pressure rather
than fixed volume; the appropriate modifications of the MFE approach
are described in~\cite{SolWarCat01}.

\section{Results}
\label{sec:results}

%%%%%%%%%%%%%%%%%%%%%%%%%%%%%%%%%%%%%%%%%%%%%%%%%%%%%%%%%%%%%%%%%%%%%%
%Present phase diagram for Schulz distribution of u (T on vertical
%axis?); cutoff is u=4? (l=2). Monodisperse limit $T_c \approx 0.220$,
%$1/T_c \approx 4.54$. Show phase diagram as is, with inset in terms of
%$T$ and whole range of $\delta$? Quadratic widening of gap with
%$\delta$ as Sluckin predicted; onset of N-I is only perturbed very
%little. Then dramatic broadening towards
%$\delta=57\%$ polydispersity in $u$.  Later more and more phases -
%though could well be preempted by smectic, crystal etc. Small range of
%$\delta$ with re-entrance.

We concentrate on the example case of a Schulz parent distribution of
interaction strengths, given by
\be
\parent \propto u^z e^{-(z+1)u}
\ee
The coefficient in the exponent has been chosen such that the mean
interaction strength $\langle u \rangle_0$ is fixed to unity, thus
setting our energy and temperature scale. The width of the
distribution is characterized by its standard deviation divided by its
mean. This quantity, denoted $\delta$ below, is often also referred to
simply as the polydispersity and is related to $z$ by $\delta =
(1+z)^{-1/2}$. To avoid interaction strengths which are arbitrarily
large compared to the mean, we impose an upper cutoff of $u_{\rm max}
= 4$. This affects the mean value of $u$ only negligibly in the range
of $\delta$ of interest to us, while the relation between $\delta$ and
$z$ has to be calculated numerically.

\begin{figure}
\begin{center}
\includegraphics[width=8cm]{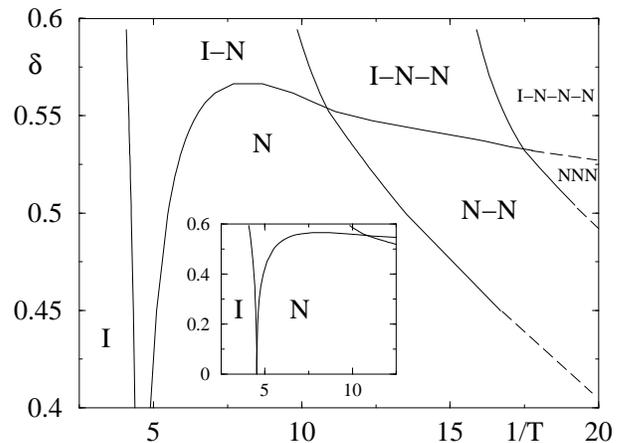}
\end{center}
\caption{Phase diagram for systems with a Schulz distribution of
interaction strengths $u$. The nature of the coexisting isotropic (I)
and nematic (N) phases is shown as a function of the inverse
temperature and the parent polydispersity $\delta$. Dashed lines
indicate the expected continuation of the phase boundaries where they
cannot be determined reliably from the numerical calculations. The
inset shows the whole range of polydispersities studied, including the
monodisperse limit $\delta\to 0$.
\label{fig:phase_diagram}
}
\end{figure}
Fig.~\ref{fig:phase_diagram} shows the resulting phase diagram,
indicating for each given parent polydispersity $\delta$ (in the range
$\delta=0\ldots 0.60$) and temperature the nature of the phases
coexisting at equilibrium. Focussing first on the inset, which shows
the whole range of $\delta$, we see that in the monodisperse limit
$\delta\to 0$ there is no coexistence gap, with the transition from
the isotropic (I) to the nematic (N) phase taking place at the
well-known Maier-Saupe temperature of $1/T_{\rm c} \approx 4.54$. At
nonzero $\delta$ an I--N coexistence region appears. Its width
initially grows quadratically with $\delta$, as predicted by
perturbation theory for small $\delta$~\cite{Sluckin89}. However, as
$\delta$ increases, the two boundaries of this I--N region are
affected in an increasingly asymmetric way. The onset temperature of
nematic ordering coming from high $T$ changes rather little,
increasing by only around 12\% as the polydispersity increases from
zero to $\delta=0.6$. The {\em lower} boundary of the I--N region, on
the other hand, is shifted to significantly lower temperatures. In
fact the broadening is so strong that the phase boundary becomes
re-entrant: for $0.553<\delta<0.567$ one has the phase split sequence
I $\to$ I--N $\to$ N $\to$ I--N $\to\ldots$ as temperature is lowered.

For polydispersities above $\delta\approx 0.45$,
Fig.~\ref{fig:phase_diagram} shows in addition that the system can
fractionate from a single-phase nematic into two and later even three
coexisting nematics as $T$ is lowered. The boundary of the N--N region
meets that of the I--N region, and at that point a three-phase
coexistence region (I--N--N) opens up where one isotropic and two
fractionated nematic phases coexist. The structure of the phase
diagram in this area is similar to the high-density phase diagram of
polydisperse hard spheres~\cite{FasSol03,FasSol04}, where two or more
fractionated solids can occur with our without a coexisting fluid
phase.

\begin{figure}
\begin{center}
\includegraphics[width=8cm]{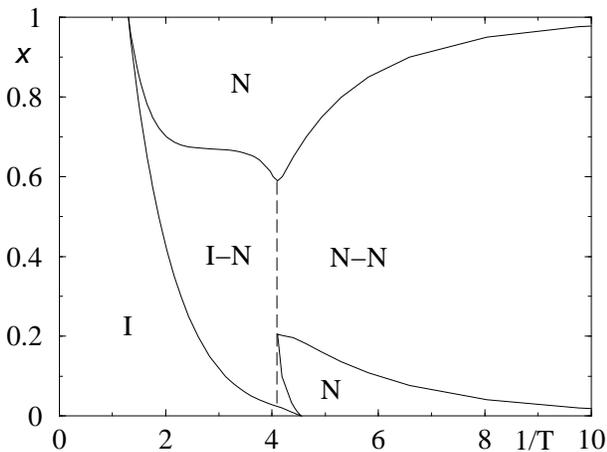}
\end{center}
\caption{Phase diagram for a bidisperse mixture with interaction
strengths $u_1=1$ and $u_2=3.5$; $x$ is the fraction of particles of
the second kind in the parent. The dashed line indicates the
temperature where three phases (I--N--N) can coexist.
\label{fig:bidisperse}
}
\end{figure}
It is worth emphasizing that all coexistence regions in the phase
diagram are caused by polydispersity; the monodisperse system only has
an I--N transition at a single temperature. Also, the presence of an
effectively infinite number of different particle species,
characterized by their respective interaction strengths $u$, means
that the number of possible phases can be in principle arbitrary. This
should be contrasted with the case of a binary
mixture~\cite{PalBerDebDun84}, which is illustrated in
Fig.~\ref{fig:bidisperse} for particles with interaction strengths
$u_1=1$ and $u_2=3.5$; $x$ is the fraction of particles of the second
kind in the parent. A region of N--N demixing again occurs at low
temperatures, but I--N--N coexistence is possible only at a single
temperature.

%%%%%%%%%%%%%%%%%%%%%%%%%%%%%%%%%%%%%%%%%%%%%%%%%%%%%%%%%%%%%%%%%%%%%%
%Demixing is driven by fractionation. Should show an example of this,
%either I-N-N or N-N; possibly I-phase is again ``spread out''?

\begin{figure}
\begin{center}
\includegraphics[width=8cm]{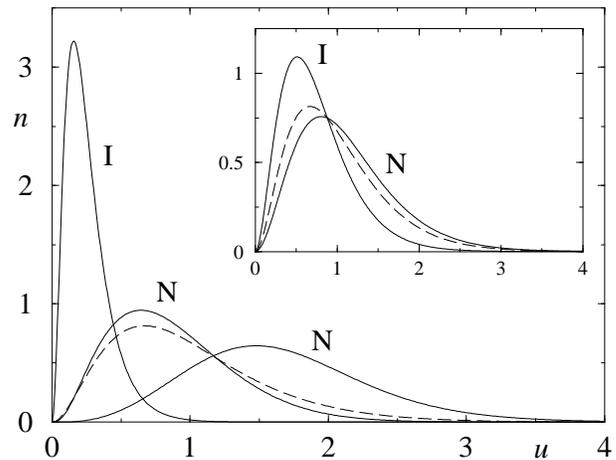}
\end{center}
\caption{Distributions of particles according to interaction strength,
$n(u)$. The solid lines show the three coexisting phases (I--N--N) at
$\delta=0.571$ and $1/T=15$; the dashed line is the parent
distribution. Inset: Same for the I--N coexistence at $\delta=0.571$
and $1/T=5.5$.
\label{fig:distributions}
}
\end{figure}
The rich phase behaviour that occurs in the polydisperse system is
driven by fractionation, i.e.\ by the fact that particles of different
interaction strength $u$ partition themselves unevenly among
coexisting phases. As an example, we show in
Fig.~\ref{fig:distributions} the distributions $\na(u)$ for the three
coexisting phases (I--N--N) into which a Schulz parent with
$\delta=0.571$ splits at $1/T=15$ (compare
Fig.~\ref{fig:phase_diagram}). The more weakly interacting particles
are found predominantly in the isotropic phase, while the particles
with larger $u$ are partititioned across the two nematic phases. This
tendency is observed more generally, as the case of two-phase I--N
coexistence in the inset of Fig.~\ref{fig:distributions}
shows. Physically, it is explained easily from the form of the
Maier-Saupe free energy: the system can lower its excess free energy
(only) by ordering nematically. This effect is stronger for the more
strongly interacting particles, which are therefore found
preferentially in the nematic phases.

\begin{figure}
\begin{center}
\includegraphics[width=8cm]{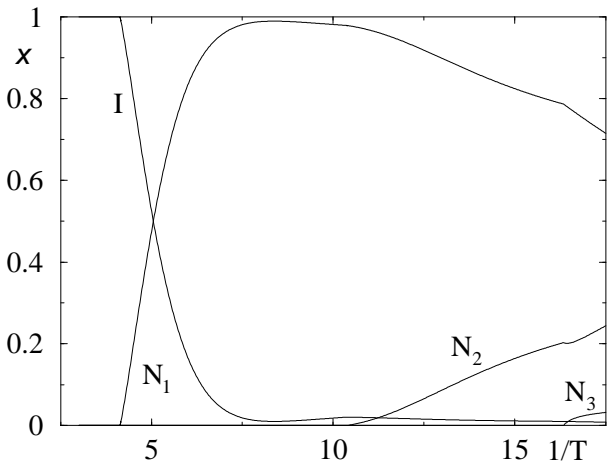}
\end{center}
\caption{Volume fractions $\xa$ of the various coexisting phases
plotted as a function of inverse temperature, for a Schulz parent with
polydispersity $\delta=0.571$.
\label{fig:volume_fractions}
}
\end{figure}
As a consequence of fractionation, the overall particle number
fractions $\xa$ found in the various coexisting phases also do not
vary linearly across coexistence
region. Fig.~\ref{fig:volume_fractions} illustrates this, for a Schulz
parent with the same polydispersity $\delta=0.571$ as in the two
examples in Fig.~\ref{fig:distributions}. One notes in particular the
non-monotonic variation of the fraction of particles in the isotropic
phase before the second nematic phase appears. This is due to the
proximity to the re-entrant I--N phase boundary in the phase diagram.

\section{Discussion}
\label{sec:conclusions}

%%%%%%%%%%%%%%%%%%%%%%%%%%%%%%%%%%%%%%%%%%%%%%%%%%%%%%%%%%%%%%%%%%%%%%
%Probably show bidisperse for comparison. Experimental situation - is
%polydisperse more favourable to seeing N-N than bidisperse, i.e. do I
%need smaller spread of u's (not clear - for bidisperse, see N-N
%connect for the first time between q=3.4 and 3.5) and/or do I get more
%reasonable T's (no, they're about a factor of 2-4 smaller than Tc; for
%bidisperse case get N-N around $T_c$ of the component that orders
%later, but again that's a factor $\approx 3$ below the $T_c$ of the
%other component)?

We have studied the effect of polydispersity in thermotropic liquid
crystals, using a Maier-Saupe theory with factorized interactions. For
a spread of interaction strengths $u$ of around $\delta=0.50$ of the
mean and higher, polydispersity causes new phase splits to appear in
the phase diagram. In particular, the theory predicts coexistence of
two or more nematics with or without an additional isotropic
phase. Together with this, a strong broadening of the I--N coexistence
region is observed, to the extent that the phase boundary between the
N and I--N regions becomes re-entrant within a small range of
$\delta$. Our calculation only considered isotropic and nematic phases
and so at least some of phase behaviour we found at lower $T$ will be
only metastable, the stable phases being e.g.\ smectic or
crystalline. Nevertheless, one would expect to be able to observe at
least the strong widening of the I--N coexistence region in
appropriate experimental systems.

From a broader theoretical point of view, our results demonstrate that
polydispersity in the attractive (and orientation-dependent)
interaction between liquid crystal particles can be sufficient to
cause demixing into two or more nematic phases. Polydispersity in
short-range repulsions as caused e.g.\ by a distribution of particle
shapes is not necessary, although from the results for binary mixtures
in the Onsager model one would expect it to reinforce the tendency
towards nematic demixing. This issue would merit further study but
promises to be very challenging. Indeed, even in the polydisperse
Onsager model without added attractive interactions it remains an open
problem to establish the existence of nematic-nematic demixing at high
densities.

One may ask how crucial the assumption of factorized interactions
$u(l,l')=u^{1/2}(l,l) u^{1/2}(l',l')$ is for the phase behaviour that
we have found. The situation is easiest to understand at low temperatures,
where one has full nematic order ($S(l)\approx 1$) and the ideal
(entropic) part of the free energy can be neglected. The system then
has to minimize the excess free energy $-\half\int\!  dl\,dl'\,n(l)
n(l')u(l,l')$. To understand whether phase separation will occur, take
first the simple case of a binary mixture, with a fraction $x$ of
particles of length (polarizability, etc) $l$ and fraction $(1-x)$ of
particles of length $l'$. The demixed state of two phases, each
containing only one type of particles, has a lower free energy than
the mixed phase if
\bea
-\half [x u(l,l) + (1-x) u(l',l')] < 
-\half [x^2 u(l,l) \\
{}+{} (1-x)^2 u(l',l') + 2x(1-x) u(l,l')] 
\nonumber
\eea
or equivalently
\be
u(l,l') < \half [u(l,l)+u(l',l')]
\label{u_convex}
\ee
One can easily show that this condition in fact guarantees that {\em
any} mixture of a finite number of components will separate into
phases each containing only one component. In a fully polydisperse
system, one then effects a cascade of demixing transitions with an
ever-increasing number of phases as $T$ is lowered, as is found in
e.g.\ models of polydisperse copolymer blends~\cite{SolWarCat01}. Our
observation of demixing into multiple nematic phases should therefore
be generic as long as\eq{u_convex} holds, i.e.\ as long as the
attractive interaction strength between different particle species is
always lower than the mean of the intra-species interaction strengths.

In future work, one could contemplate extending the present study by
accounting for the effects of density variations. To a first
approximation this can be done by dividing the interaction strengths
$u(l,l')$ by the average particle volume in the given
phase~\cite{HumLuc73}. In the polydisperse setting, this would give an
extra factor of $1/\bar{v}$ in the excess free energy, where
$\bar{v}=\int\! dl\,n(l)v(l)$ and $v(l)$ is the particle volume as a
function of $l$. One then still has a truncatable free energy, but now
with an excess part depending on two moments, $\m$ and
$\bar{v}$. Also of interest would be an extension to liquid crystal
polymers, where perturbative results for small polydispersity $\delta$
already exist~\cite{Semenov93} and there is experimental evidence for
N--N phase separation, at least in binary mixtures~\cite{CasVeyFin82}.

\bibliography{/home/psollich/references/references}
\bibliographystyle{prsty}

\end{document}